\documentclass[aps,prl,superscriptaddress,reprint]{revtex4-1}
\usepackage[dvips]{graphicx}
\usepackage{xcolor}
\begin{document}

\title{Bose-Einstein condensation of magnons in atomic hydrogen gas}

\author{O. Vainio}
\email{otarva@utu.fi}
\author{J. Ahokas}
\author{J. J\"{a}rvinen}
\author{L. Lehtonen}
\author{S. Novotny}
\author{S. Sheludiakov}
\author{K.-A. Suominen}
\author{S. Vasiliev}
\affiliation{Department of Physics and Astronomy, University of Turku, 20014 Turku, Finland}
\author{D. Zvezdov}
\affiliation{Department of Physics and Astronomy, University of Turku, 20014 Turku, Finland}
\affiliation{Kazan Federal University, 420008, 18 Kremlyovskaya St, Kazan, Russia.}
\author{V. V. Khmelenko}
\author{D. M. Lee}
\affiliation{Department of Physics and Astronomy and Institute for Quantum Science and Engineering, Texas A$\&$M University, College Station, TX, 77843, USA}

\begin{abstract}
We report on experimental observation of BEC-like behaviour of quantized electron spin waves (magnons) in a dense gas of spin polarized atomic hydrogen. The magnons are trapped and controlled with inhomogeneous magnetic fields, and described by a Schr{\"o}dinger-like wave equation, in analogy to the BEC experiments with neutral atoms. We have observed the appearance of a sharp feature in the ESR spectrum displaced from the normal spin wave spectrum. We believe that this observation corresponds to a sudden growth of the ground state population of the magnons and emergence of their spontaneous coherence for hydrogen gas densities exceeding a critical value, dependent on the trapping potential. We interpret the results as a BEC of non-equilibrium magnons which were formed by applying the rf power.
\end{abstract}


\maketitle 

In contrast to the Bose-Einstein condensation (BEC) experiments with real particles, e.g.\ with alkali atoms \cite{Pethick2008}, a condensation of quasiparticles cannot be achieved by lowering the temperature of a thermalized system, but instead a non-equilibrium state is necessary \cite{Safonov2013}. This overpopulation of quantum states compared to the thermal equilibrium population given by the Planck distribution is achieved by injecting additional quasiparticles to the system externally, i.e.\ pumping the system. BEC-like behavior, or spontaneous coherence, in systems of coupled oscillators has been predicted by H. Fr{\"o}hlich \cite{Frohlich1968} and is often referred to as the Fr{\"o}hlich coherence. In recent years a Bose-Einstein-like condensation of quasiparticles has been reported in several distinct systems. These include exciton polaritons \cite{Kasprzak2006}, triplet states in magnetic insulators \cite{Ruegg2003}, magnons in ferromagnets \cite{Demokritov2006a} and liquid $^3$He \cite{Bunkov2010c}, and photons in a microcavity \cite{Klaers2010}. Understanding the properties of quasiparticles is increasingly important due to advancing technologies pushing ever further into the quantum realm.

Quantized electron spin waves (magnons) form a quasiparticle system in a dense quantum gas of ultracold atomic hydrogen \cite{Vainio2012}. The definition of a quantum gas is that the thermal de Broglie wavelength $\Lambda_{th}$ substantially exceeds the scattering length of elastic collisions $a_s$. By tuning the magnetic field profile in our experiment we can modify the magnon-trapping potential. Furthermore, by changing the atomic hydrogen gas density we are able to modify the spin-exchange interaction strength and thereby the dynamics of the magnons. These two tools together provide unique control of the quasiparticle dynamics and allow the formation of a BEC of magnons similar to that observed in the BEC of neutral atoms \cite{Pethick2008}.

In the present study of dense H gas by electron spin resonance (ESR), we observed a sudden change in the spectrum of the trapped magnons: a sharp and intense peak corresponding to their ground state in the trap grows rapidly after H gas density exceeds a critical value. We believe that this is associated with the emergence of a spontaneous coherence in the system. Based on these two observations we conclude that the magnons undergo a transition to a BEC when their ground state energy becomes equal to the chemical potential.

The origin of spin waves in quantum gases is fundamentally different from that in ferromagnets, where the phenomenon is due to the strong \textit{electron} exchange interaction \cite{Keffer1966}. In quantum gases weak exchange effects of identical \textit{atoms} lead to a phenomenon known as the identical spin rotation (ISR) effect \cite{Lhuillier1982,Lhuillier1982a}. In the ISR effect, the spins of the interacting atoms rotate around the sum of the spins during binary collisions. The macroscopic propagation of a spin perturbation, the spin wave, is a cumulative result of numerous ISR collisions. The dynamics of the transverse magnetization $S_+ = S_x + iS_y$ due to the ISR effect is described by the complex spin transport equation, which for strong magnetic fields and small $S_+$ is \cite{Lhuillier1982,Lhuillier1982a,Levy1984}

\begin{equation}
i \frac{\partial S_{+}}{\partial t}= D_0 \frac {\varepsilon}{\mu^*} \nabla^{2} S_{+} + \gamma \delta B_0 S_{+}, \label{ISR equation}
\end{equation}

\noindent where $D_0$ is the spin diffusion coefficient in the unpolarized gas, $\varepsilon = \pm 1$ for bosons/fermions, $\gamma$ is the gyromagnetic ratio, $\delta B_0$ is the deviation of the magnetic field from its average value $B_0$ and $\mu^* \propto \Lambda_{th} / a_s$ is the spin wave quality factor. Writing $m^*=-\hbar\mu^*/2D_0\varepsilon$ equation (1) becomes a Schr{\"o}dinger equation for a particle with an effective mass $m^*$. The effective mass of the magnons and the trapped magnon ground state energies $\epsilon_{0}$ both depend on the hydrogen gas density $n_H$, since $D_0\propto n_H^{-1}$ \cite{Bigelow1989,Levy1984}. For our experimental conditions, the effective mass of the magnons is on the order of the free electron mass, making the observation of BEC-like behavior plausible at temperatures substantially higher than those required for the BEC of H atoms \cite{Fried1998}.

\begin{figure*}
\includegraphics[width=17 cm]{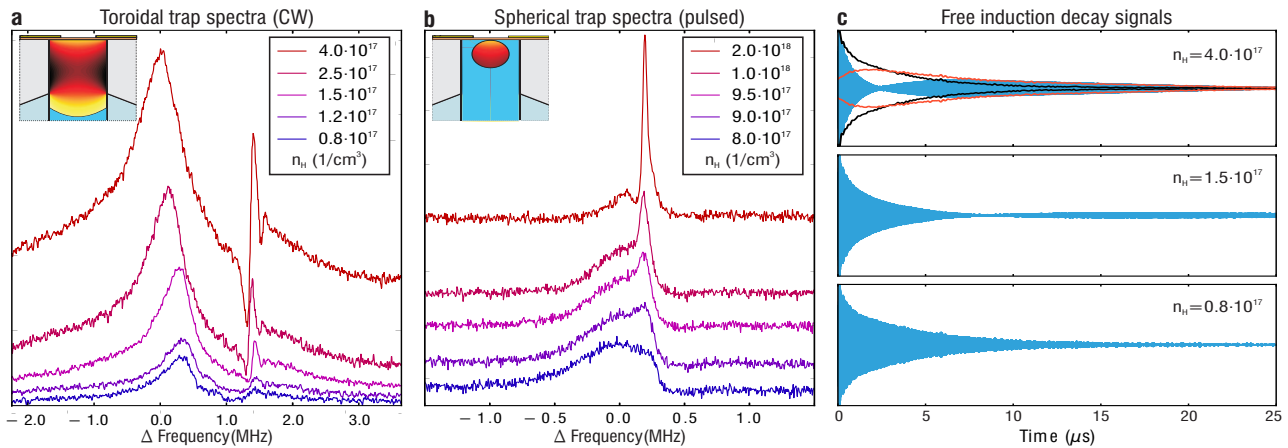}
\caption{CW (\textbf{a}) and pulsed (\textbf{b}) ESR spectra with signal from trapped magnons at five different hydrogen gas densities in the two different trapping geometries (depicted in inserts) described in \cite{SM1}. In the low density spectra in \textbf{a} the left side peak is due to ESR absorption at the rf field maximum and the right side peak is signal from the trapped magnons. Due to broadening the peaks are partially superimposed in the higher density spectra. In \textbf{b} only the signal from trapped magnons is visible. In both trapping geometries the emergence of the spontaneous coherence is visible as a narrow high-amplitude feature. \textbf{c}: Pulsed ESR (FID) signals below (bottom), above (top), and approximately at the critical hydrogen gas density (middle). In the top panel the envelopes of the decomposition of the measured signal (blue) into the coherent (red) and incoherent (black) signals is shown. The node in the blue signal is due to destructive interference between coherent and incoherent parts. The continuing growth of the coherent signal amplitude after the end of the excitation pulse is a clear sign of the spontaneity of the coherence. The FIDs are recorded in the toroidal geometry.}\label{Spectra_and_FIDs}
\end{figure*}

For electron spin waves in hydrogen gas, regions of high magnetic field are minima of magnetic potential, which is much stronger and of opposite sign than that for nuclear spin waves \cite{Bigelow1989}. Due to weakness of the ISR effect, the influence of the magnetic field inhomogeneities on the dynamics of electron spin waves in hydrogen gas is much stronger than that for ferromagnets.

The experiments are conducted in a high magnetic field of 4.6 T, and therefore the electron and nuclear spin projections, S and I, are good quantum numbers. For the same reason the induced transverse magnetization $S_+$ is a small perturbation to the spin polarization. Most of the other spin wave experiments in quantum gases have been conducted in low fields, where the total spin F is a good quantum number \cite{Lewandowski2002,Deutsch2010a}. In these experiments spin self-rephasing and spatial segregation have been observed, which is consistent with the onset of Bose-Einstein condensation. In this work we report the first observation of spontaneous coherence due to BEC-like behaviour of magnons in a quantum gas.

For excitation (pumping) and detection of the spin waves we use ESR spectroscopy at $129$ GHz resonant with the ESR transition from the electron-nuclear polarized hyperfine state \cite{Silvera1986}. For generating the spin waves, a spatially inhomogeneous excitation is created by an evanescent tail of the rf field \cite{Vainio2012}. We use both continuous wave (CW) and pulsed ESR methods. The temperature in our experiments ranges from 200 mK to 600 mK. Accurate control of the gas density and volume is achieved by a liquid helium piston operated with a fountain valve, which also provides an independent measurement of the hydrogen gas pressure \cite{Vainio2012,Tommila1987,SM1}. By compressing the gas with this method we are able to produce hydrogen gas densities from $10^{16}$ cm$^{-3}$ to $5 \cdot 10^{18}$ cm$^{-3}$. The experimental setup is described in more detail in \cite{SM1}.

The inner walls of the hydrogen gas volume are covered with liquid helium and function as reflective mirrors for the spin waves \cite{Bigelow1989}. In addition, there are magnetic field inhomogeneities inside the sample volume, which in combination with the liquid helium walls create a potential well for the electron spin waves. We can control the strength and position of these magnetic field deviations with sets of gradient coils outside the sample volume. Additionally, due to the behaviour of the liquid helium piston, we produced two distinctively different magnon trap geometries: one toroidal and another with the shape of a spherical cap \cite{SM1}. In the toroidal trap $n_H$ and $\epsilon_0$ are lower and the density of magnon states is higher than in the spherical trap.

In Fig.\ 1(a and b) several ESR spectra recorded both below and above the critical hydrogen gas density in both trapping geometries are shown. A  large number of unresolved spin wave modes trapped in the potential minimum created by magnetic field inhomogeneity are seen as a broad peak on the right side of the ESR spectrum \cite{Vainio2012}. Above the critical hydrogen gas density a high-amplitude, narrow-width spectral feature emerges from the trapped magnon peak. Notice the difference in phase between this trapped magnon signal and the main ESR absorption line, which will be discussed later. 

We interpret this narrow spectral feature as the emergence of spontaneous coherence resulting from the redistribution of the spin wave modes to the lowest energy mode in the magnetic potential well. Our reasoning is based on the following five arguments: 1) In all our observations, the position of this feature corresponds to the position of the bottom of the magnon potential. 2) The coherence is established after a short build-up time, not directly at the time of excitation, shown in Fig.\ 1(c). 3) We observe critical behavior relative to the hydrogen gas density. 4) We observe spin transport from higher to lower potential. 5) The coherence has a consistent phase, the offset of which depends on the spatial separation between the trapped magnons and the excitation region.

The coherence is seen in both CW and pulsed ESR spectra, with small variations in appearance due to the differences in the detection techniques. In CW measurements the magnetic field offset is swept and the exciting rf field is kept constant. Due to magnetic field inhomogeneities the spectrum recorded with a CW field sweep is a map of the local absorption (and dispersion) rates inside the sample volume.

In the CW ESR, excitation and detection happen simultaneously and always at the prevailing resonance conditions. Evidence for the coherence in the CW measurements is indicated by the  extremely small line width associated with the with the trapped magnons. Another observation confirming the coherent nature of the ground state signal is its consistent phase difference as compared to the ESR signal from the hydrogen atoms located at the top of the sample volume, in the rf maximum \cite{SM1}. These atoms produce a broad peak at the on the left side of the spectra in Fig.\ 1a. The phase difference is in accordance with the magnetic field maximum's spatial separation from the main excitation region. For example, an estimated $l = 0.25$ mm separation between the magnetic field maximum and the rf field ($\lambda = 2.34$ mm) coupling orifice results in a phase difference $\Delta \varphi = 4\pi l/\lambda \approx 1.34$, in good agreement with an observed $\Delta \varphi$ of approximately $\pi/2$ (see Fig.\ 1a) \cite{SM1}. 

In the pulsed ESR technique the magnetic field is kept constant and short pulses of rf excitations are sent to the sample volume. Between the excitation pulses the free induction decay (FID) signal is recorded. The pulsed spectra in the frequency domain are Fourier transforms of the FID signals, usually averaged over several thousand pulse-detection periods.

In Fig.\ 1(c) FID signals below, above, and approximately at the critical density are shown. The FID signal with the coherence has a node in its envelope due to destructive interference between the coherent and incoherent signals. In the top panel of Fig.\ 1(c) the decomposition of the measured signal into coherent and incoherent parts is shown. The amplitude of the coherent part is still growing after the excitation pulse has been switched off. This is a clear sign of spontaneous reorganization of the spins within the sample instead of induced coherence by the excitation pulse. The incoherent signal decays rapidly because of differences in oscillation frequencies of the large number of magnon modes out of which it is composed. The coherent signal, in contrast, originates from a single (ground state) mode and therefore has a longer decay time. The coherence has been observed in various potentials in two distinct sample gas geometries \cite{SM1}. Similar signals have also been seen in systems of liquid $^3$He \cite{Bunkov2010c} and a quantum gas of $^{87}$Rb \cite{Deutsch2010a}.

\begin{figure}
\includegraphics[width=8 cm]{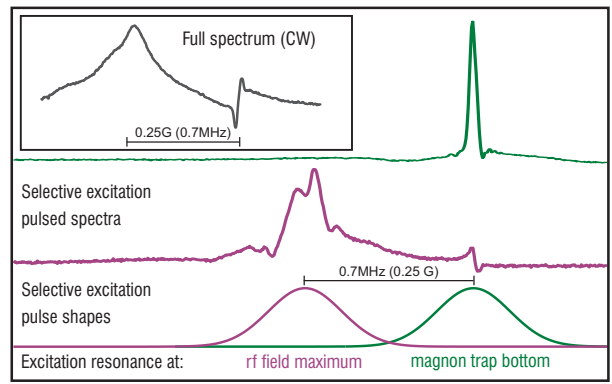}
\caption{Spectra measured with selective excitation pulses. Green and magenta curves show selective excitation pulse shapes and the measured spectra correspondingly. Excitation in resonance with hydrogen atoms located at the rf field maximum (purple curves), spatially separated from the minimum of the magnon potential, still produces a signal from the bottom of the potential. Similar excitation in resonance with the bottom of the potential (green curves) only produces signal from the potential minimum. Inset shows CW spectrum.}\label{FID_comparison}
\end{figure}

By varying the pulse length and shape, the spectral width of the excitation can be manipulated. This can be used for selective excitation of limited regions of the sample volume. Fig.\ 2 shows two different pulsed ESR spectra resulting from two different locations being selectively excited. An excitation that is resonant at the top of the sample volume, at a distance from the magnon trap bottom, still produces a disproportionately strong signal from the magnon trap bottom (purple curves). In contrast, an excitation resonant at the trap bottom only produces a single narrow-bandwidth high-amplitude spectral peak (green curves). We interpret this as spin transport from higher potential towards the bottom of the spin wave trap.

During hydrogen atom gas compressions, the spectra evolve smoothly until a sudden growth of the high-amplitude coherent signal begins after the critical density has been reached. The observed $n_H$-dependent criticality in the formation of the coherence in different potentials follows the behavior of the simulated ground state energies $\epsilon_0(n_H)$ for trapped magnons. We argue that changes in the ground state energy, rather than in the chemical potential, will lead to the ground state overpopulation. In Fig.\ 3 curves corresponding to critical behavior of the coherent peak amplitudes of trapped magnon signals for three different potentials are plotted against the hydrogen gas density. In the same figure the simulated behavior of the corresponding ground state energies are also shown. The ground state energies corresponding to our experimental conditions have been found by numerically calculating the eigenmodes and frequencies of the magnon modes for several different hydrogen gas densities using equation (1). The density dependence of the ground state energy closely follows the relation $\epsilon_0 \propto n_H^{-1/2}$.

As with other quasiparticle systems the chemical potential is non-zero only under pumping, e.g.\ while introducing out-of-equilibrium magnons into the system. In particular, the Bose-distribution

\begin{equation}
n_k = \left[e^{\frac{\epsilon_k-\mu}{k_B T}}-1\right]^{-1}
\label{Bose_distribution}
\end{equation}

\noindent is only valid for the quasiparticles while the system is being pumped and the inter-state relaxation rate is much faster than the relaxation to the thermal bath. The details of the magnon thermalization are still unknown. We assume that the rate of thermalization is fast enough to establish the distribution given by Eq. \ref{Bose_distribution}.  

For ISR magnons in atomic hydrogen gas the magnon pumping rate also depends on the hydrogen gas density as $I_p \propto W n_H$, where the probability $W$ to create a magnon is proportional to the rf power. Consequently, with a fixed rf power, the pumped magnon density is directly proportional to $n_H$. The magnon density could also be increased by increasing the rf power, but for spin polarized hydrogen this approach leads to rapid recombination. In the high temperature limit $k_B T \gg \epsilon_0$, well justified in our experiments, the chemical potential $\mu$ depends on the pumping rate $I_p$, temperature $T$, ground state energy $\epsilon_0$ and the dissipation rate to the thermal bath $\tau_1^{-1}$ as \cite{Bugrij2008}

\begin{equation}
\mu = I_p \tau_1 \frac{\epsilon_0^2}{k_B T}.
\label{Chemical_potential}
\end{equation}

\begin{figure}
\includegraphics[width=8 cm]{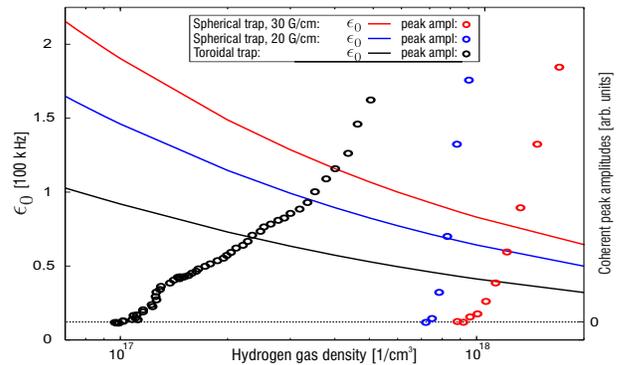}
\caption{Ground state energies for trapped magnons and coherent peak amplitudes of trapped magnon signals vs. hydrogen atom density for three different potentials. Solid lines show the simulated ground state energies $\epsilon_{0}$ and circles the detected coherent signal amplitudes vs. hydrogen atom density. Black: toroidal trap, no applied linear gradient. Blue: spherical trap with $\approx 20$ G/cm gradient. Red: spherical trap with $\approx 30$ G/cm gradient. A clear density- and potential-dependent criticality is observed in the appearance of the coherent signal.}
\label{Critical_density_comparison}
\end{figure}

From equation (3) it follows that for a given potential and fixed rf power the induced chemical potential does not depend on $n_H$, because the density-related changes in the pumping rate compensate for the changes in the ground state energy. Therefore the magnon ground state population $n_0$ in equation (2) depends on density only via the ground state energy $\epsilon_0$.

According to equation (2) the divergence of the ground state population $n_0$ corresponding to BEC occurs when $\epsilon_0 = \mu$ and therefore the equation for the critical pumping rate is 

\begin{equation}
I_{pc} \tau_1 = \frac{k_B T}{\epsilon_{0c}}.
\label{Coherence_condition}
\end{equation}

The observed difference in the critical densities between the potentials is explained by two separate but related factors. Firstly, the ground state energies for the potentials differ considerably. Secondly, due to significant differences in the sample gas geometry, the magnon pumping rates $I_p$ are different \cite{SM1}.

In this work, we have demonstrated that in a quantum gas of atomic hydrogen, a BEC of magnons can be achieved. The condensed magnons are trapped in a potential well created by a local maximum of the magnetic field and the walls of the experimental cell. The phenomenon is a direct consequence of the Bose statistics, as pumped bosonic quasiparticles preferably accumulate in the ground state \cite{Bugrij2008}. Contrary to the customary behaviour of the atomic (real particle) Bose-Einstein condensates \cite{Fried1998}, the magnons condense at a lower hydrogen gas density in the potential with lower ground state energy and higher density of states. This follows from the relation $n_0 \propto (\epsilon_0 - \mu)^{-1}$ ($\mu \approx$ constant, but $ \epsilon_0 \propto n_H^{-1/2}$) for the ground state population. 

The ability to modify the trapping potential by precisely tuning the magnetic field profile opens up many possibilities for further experimentation. An interesting future experiment would be to prepare two tunable coupled traps and study interference effects between two magnon condensates, in analogy to the early experiments with $^3$He \cite{Borovik-Romanov1988} and to the neutral atom interference \cite{Andrews1997}. Also, the details of the thermalization of magnons in hydrogen gas represent a very interesting topic for a special study. 


\begin{acknowledgments}
This work was supported by the Academy of Finland (Grants No. 268745, and 260531), the Wihuri Foundation and NSF grant No. DMR 1209255. We thank G. Volovik, M. Krusius, V. Eltsov, Yu. Bunkov and N. Bigelow for useful discussions.
\end{acknowledgments}


%

\pagebreak
\widetext
\begin{center}
\textbf{\large Supplemental material: Bose-Einstein condensation of magnons in atomic hydrogen gas}
\end{center}
\setcounter{equation}{0}
\setcounter{figure}{0}
\setcounter{table}{0}
\setcounter{page}{1}
\makeatletter

\subsection*{Hydrogen gas volume and compression technique}

The size and shape of the hydrogen gas volume in the experimental cell we utilized for these experiments can be controlled with a liquid helium piston. The liquid helium piston is operated by a fountain valve, i.e. a super-leak between the hydrogen gas volume (sample volume) and a liquid helium container. The liquid helium container is outfitted with capacitor plates which, in conjunction with an inductor and tunnel diode oscillator, provide accurate measure of the liquid helium level inside the container. By tuning the temperature difference between the container and the hydrogen gas volume we are able to pump the liquid helium between the two volumes. Each experimental cycle consists of first filling the hydrogen gas volume with hydrogen followed by compression with the liquid helium piston. A typical experimental cycle lasts a few hundred seconds, during which we record several tens of spectra. In Fig.\ 1 schematic of the compression sequence is shown.

During a weak compression sequence the steady state liquid helium level stays well below the top of the small diameter cylinder and the highest hydrogen density is reached at the end of the active compression (Fig.\ 1 b). During the following stage with a constant temperature difference between the liquid helium container and the hydrogen gas volume, the hydrogen gas recombination leads to decreasing hydrogen atom number and density. These compressions will eventually end with an empty sample volume above the liquid helium surface. We can control the peak density and final geometry (length) of the cylinder. 

In the end of sufficiently strong compressions, the hydrogen gas evolves into a bubble at the top of the cylinder, fully immersed in liquid helium (Fig.\ 1 c). The lifetime of the bubble is defined by the strength of the compression, but during a carefully controlled compression we are able to reach a 60-80 second lifetime of the bubble. The hydrogen gas density in the bubble is increasing until the end of the bubble lifetime, even though the number of hydrogen atoms is decreasing, as the atomic hydrogen gas is continuously recombining inside the bubble with the molecules exiting through the helium surface. This happens because the pressure $p_{He}$ due to the liquid helium surface tension $\sigma$ is related to the radius of the bubble $r$ as $p_{He}=2\sigma/r$. Compressions where a bubble of atomic hydrogen is formed will end up as a liquid helium filled sample volume.

Also, during the bubble stage of the compression, the overlap between the hydrogen gas volume and the excitation rf field is decreasing and therefore the magnon pumping rate is also decreasing.

By varying the strength of the compression we are able to produce atomic hydrogen gas densities differing by more than two orders of magnitude, from $10^{16}$ cm$^{-3}$ to $5 \cdot 10^{18}$ cm$^{-3}$. This is important because the spin diffusion coefficient $D_0$ and the pumped magnon density depend on the atomic hydrogen gas density and therefore also the dynamics of the magnons depend on $n_H$ Most importantly, in relation to our experimental conditions, the magnon ground state energies $\epsilon_0$, depend on $n_H$.

\begin{figure}[hb]
\includegraphics[width=10 cm]{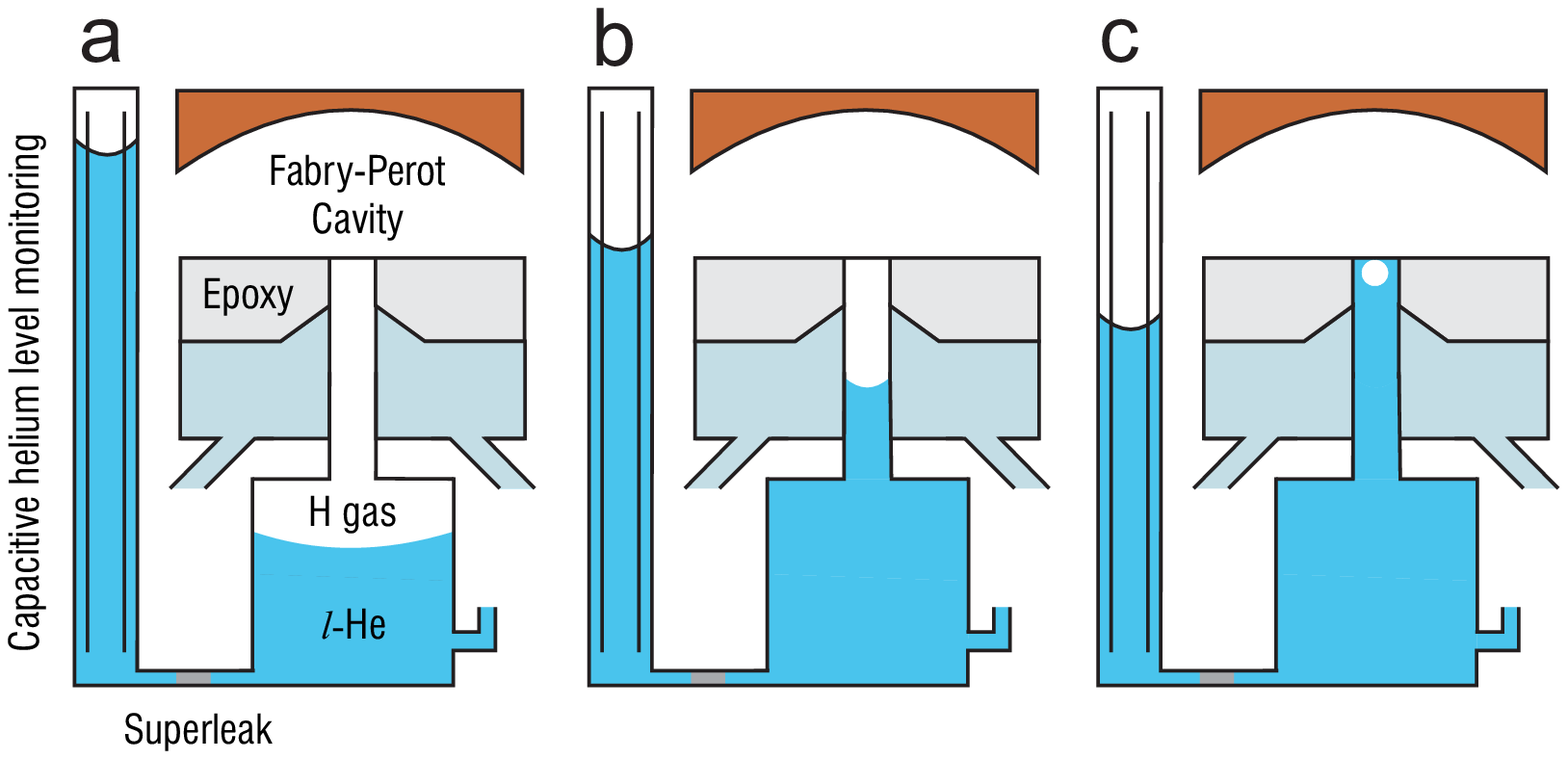}[!b]
\caption{Schematic of the hydrogen gas compression using a liquid helium piston.} \label{Compression}
\end{figure}

\subsection*{Magnon potential geometries}

The top of the hydrogen gas sample volume inside which the spin waves are detected is a 0.5 mm diameter cylinder, the length of which can be controlled with a liquid helium piston compressing the gas. The total length of the small diameter cylinder is 6 mm, but the magnetic field maxima and therefore the trapped magnons are located within 1 mm from the top of the cylinder, shown in Fig.\ 2(a and b). The evanescent rf field for detection and excitation has a characteristic length of 80 $\mu$m, shown with a dashed line. The plastic walls of the cylinder are covered with the liquid helium film and function as reflective mirrors for the spin waves. In addition to the reflective boundaries of the liquid helium walls there are magnetic field inhomogeneities inside the sample volume. The inhomogeneities are a result of stray magnetic fields due to weak magnetization of the epoxy meniscus surrounding the top of the hydrogen volume cylinder.

\begin{figure}
\includegraphics[width=10 cm]{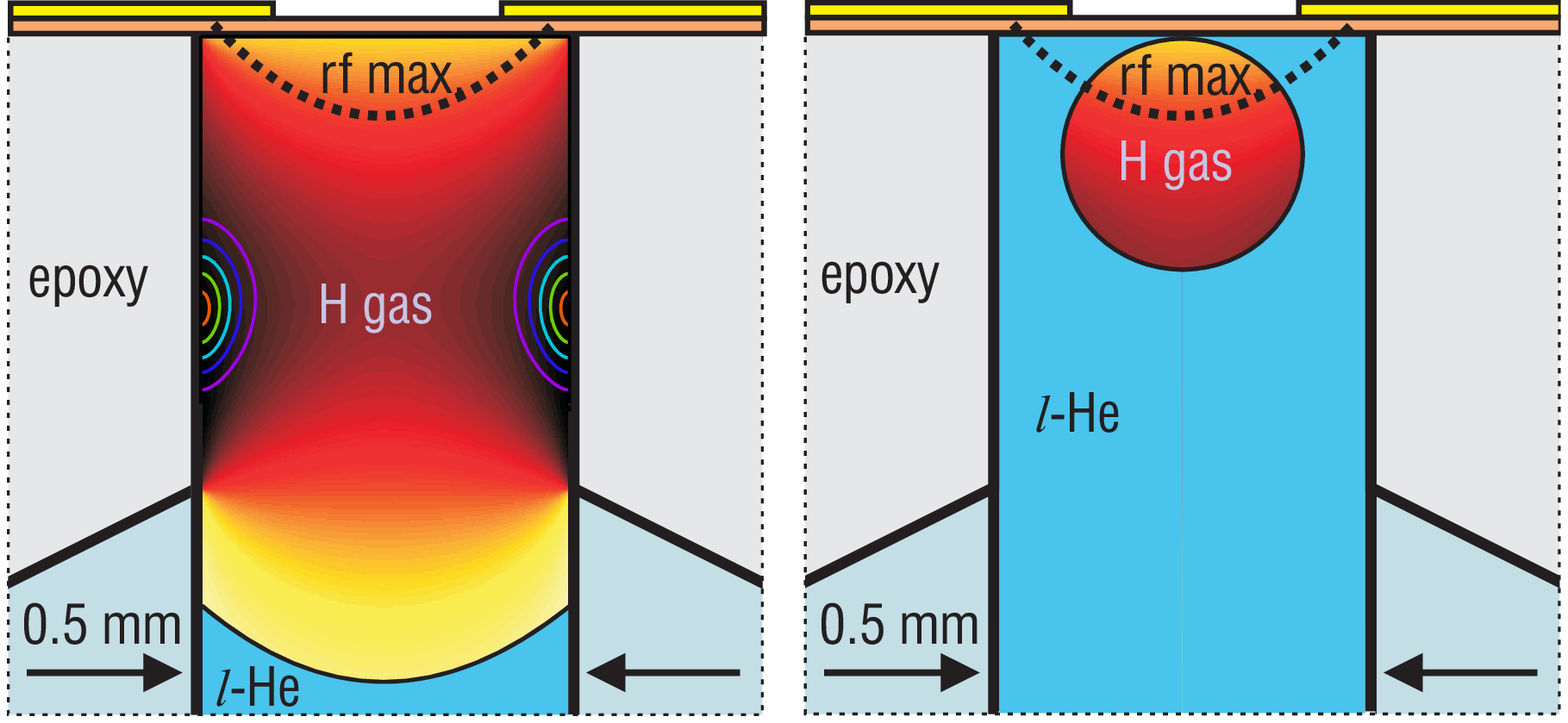}
\caption{Sectional drawings of the toroidal (\textbf{a}) and spherical (\textbf{b}) magnon traps within the atomic hydrogen gas volume. In both \textbf{a} and \textbf{b} the vertical center line is the axis of symmetry. Dark areas correspond to higher magnetic field, i.e.\ minima of the potential. An evanescent tail of the rf field from the Fabry-Perot cavity is coupled to the sample gas through a sub-critical hole in the lower cavity mirror. In \textbf{a} also the contours of the amplitude (red is largest, purple is smallest) of the numerically calculated ground state for hydrogen density $1.2\cdot 10^{17}$ cm$^{-1}$ and temperature $300$ mK are shown.} \label{Trap_geometries}
\end{figure}

We can control the strength and position of these magnetic field deviations with sets of gradient coils outside the sample volume. The magnetic field maxima are always located adjacent to a liquid helium boundary, with either the liquid helium film covering the plastic wall or the surface of the bulk liquid helium. The net effect of the liquid helium surface and the magnetic field maximum is that the potential for the magnons has a minimum at the boundary between the hydrogen gas and liquid helium. The relatively smooth potential curvature due to the magnetic field is cut off at the minimum by the reflecting, wall-like potential due to the liquid helium surface.

The shape of the toroidal magnon trap remains unchanged during the compression, even though the height of the hydrogen gas volume is radically changing, until the liquid helium level reaches the top most 0.5 mm of the hydrogen gas volume. Therefore the only variable during the sample evolution in the toroidal trap is the hydrogen gas density. In contrast, the diameter of the spherical trap is constantly changing during the time evolution of the spherical trap. Due to the additional compression provided by the liquid helium surface tension, the hydrogen gas densities in the spherical trap are approximately an order of magnitude higher than the densities in the toroidal trap.

\begin{figure}
\includegraphics[width=10 cm]{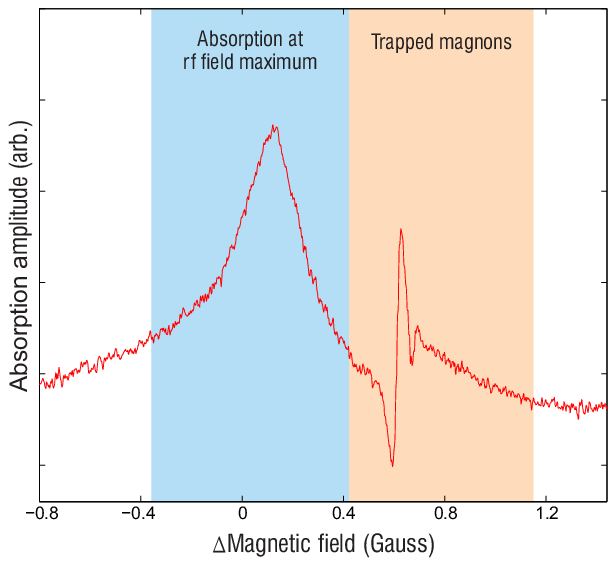}
\caption{Full ESR spectrum recorded by CW in the toroidal trapping geometry. In this gradient the absorption at the rf field maximum is separated from the absorption by the trapped magnons by approximately 0.6 Gauss. The separation between these two absorption maxima solely depends on the magnetic field difference between the magnetic field maximum and the field at the rf field maximum, and is constant for a given setting of the external gradient coils. The phase difference between the coherence and the absorption at the rf field maximum on the other hand depends on the spatial separation between the magnetic field maximum and the rf field maximum.}
\label{Toroidal_spectrum}
\end{figure}

In Fig.\ 3 a CW spectrum recorded in the toroidal trapping geometry is shown. The magnetic field gradient for this spectrum is the one depicted in Fig.\ 2a. The phase of the spectrum is adjusted according to the absorption peak originating from the rf field maximum. Compared to the peak from rf field maximum the trapped magnon signal has a $\pi/2$ phase difference. This difference is particularly well visible in the narrow feature, i.e. the coherent part of the spectrum. The phase difference is due to the spatial separation between the origins of these two signals, as explained in the main article.

\end{document}